\documentclass[useAMS,usenatbib]{mn2e}

\usepackage{graphicx}
\usepackage{scalefnt}
\usepackage{amssymb}

\title[Improved models of merger times]{An improved prescription for merger time-scales from controlled simulations}
\author[\'A. Villalobos, G. De Lucia, S. Weinmann, S. Borgani, and G. Murante]{
\'A. Villalobos$^{1}$\thanks{villalobos@oats.inaf.it}, G. De Lucia$^{1}$, S. Weinmann$^{2}$, 
S. Borgani$^{1,3,4}$, and G. Murante$^{1}$\\
$^{1}$INAF - Astronomical Observatory of Trieste, via G.B. Tiepolo 11, 34143 Trieste, Italy\\
$^{2}$Leiden Observatory, Leiden University, P.O. Box 9513, 2300 RA Leiden, The Netherlands\\
$^{3}$University of Trieste, Department of Physics, via Valerio 2, 34127 Trieste, Italy\\
$^{4}$INFN - National Institute for Nuclear Physics, via Valerio 2, 34127 Trieste, Italy
}

\begin{document}

\date{Accepted --- . Received --- ; in original form ---}

\pagerange{\pageref{firstpage}--\pageref{lastpage}} \pubyear{---}

\maketitle

\label{firstpage}

\begin{abstract}
We compare three analytical prescriptions for merger times available 
from the literature to simulations of isolated mergers. We probe three 
different redshifts, and several halo concentrations, mass ratios, 
orbital circularities and orbital energies of the satellite. We find 
that prescriptions available in the literature significantly 
under-predict long timescales for mergers at high redshift. We argue 
that these results have not been highlighted previously either because 
the evolution of halo concentration of satellite galaxies has been 
neglected (in previous isolated merger simulations), or because long 
merger times and mergers with high initial orbital circularities are 
under-represented (for prescriptions based on cosmological simulations).
Motivated by the evolution of halo concentration at fixed mass, an 
explicit dependence on redshift added as 
\mbox{$t_{\rm merger}^{\rm mod}(z) = (1+z)^{0.44} t_{\rm merger}$} 
to the prescription based on isolated mergers gives a significant 
improvement in the predicted merger times up to $\sim$20~$t_{\rm dyn}$ 
in the redshift range 0$\le$$z$$\le$2. When this modified prescription 
is used to compute galaxy stellar mass functions, we find that it leads 
up to a 25 per cent increase in the number of low mass galaxies 
surviving at $z$=0, and a 10 per cent increase for more massive galaxies. 
This worsen the known over-prediction in the number of low mass galaxies 
by hierarchical models of galaxy formation. 
\end{abstract}


\section{Introduction}

\defcitealias{boylan-kolchin2008}{B08}
\defcitealias{jiang2008}{J08}
\defcitealias{mccavana2012}{M12}

Galaxies orbiting in dense environments, such as groups or clusters,  
suffer a continuous loss of energy and angular momentum under the 
effect of dynamical friction against the medium of their parent halo. 
Because of this, galaxies on bound orbits spiral-in towards the densest 
regions of their local environment on a given timescale. This merger 
timescale $t_{\rm merger}$ is usually modelled as a function of the 
dynamical timescale of the main halo, $t_{\rm dyn}$, the galaxy-main halo 
mass ratio, and the orbital energy and circularity of the infalling galaxy. 
All of these quantities are considered at the time a galaxy crosses 
the virial radius of the main halo along its infalling orbit. Merger 
timescales are defined as the time until the galaxy has lost a significant 
fraction of either its initial mass via tidal stripping, or of its 
initial orbital angular momentum. 

Accurate estimates of how long a galaxy survives within a halo while 
being affected by dynamical friction are fundamental for theoretical 
studies of galaxy evolution. Indeed, merger times play a key role in 
the evolution of galaxies' stellar masses, morphologies, colours, and 
gas content \citep[e.g.][]{cox2008,delucia2010,delucia2011}.

In recent years, \citeauthor{boylan-kolchin2008} (\citeyear{boylan-kolchin2008}, 
\citetalias{boylan-kolchin2008}), \citeauthor{jiang2008} (\citeyear{jiang2008}, 
\citetalias{jiang2008}), and \citeauthor{mccavana2012} (\citeyear{mccavana2012}, 
\citetalias{mccavana2012}) have provided different prescriptions to 
estimate merger timescales. All three prescriptions are variations of 
the analytic description derived by \citet{chandrasekhar1943} for the 
drag force suffered by a point mass object as it moves through a uniform 
background medium of less massive particles \citep[see][]{binney-tremaine1987}. 
The aforementioned prescriptions are obtained either by simulating single 
isolated mergers at $z$=0 that probe a given parameter space, or by 
collecting a sample of mergers from cosmological simulations within a 
given redshift range. In this paper, we compare these prescriptions 
with controlled simulations of isolated galaxies being accreted onto a 
group-like halo at three redshift epochs. We explore several halo 
concentrations in both systems, merger mass ratios, and orbital parameters 
of accreted galaxies. Our main goal is to determine whether the implicit 
dependency of these prescriptions on redshift (via $t_{\rm dyn}$)
is enough to account for relevant properties of haloes that are known 
to evolve with cosmic time, such as halo concentration.  
   
The layout of this paper is as follows: Section \ref{sec-setup} describes 
the set-up of our experiments; Section \ref{sec-results} describes our 
results, comparing them to models from 
the literature; in Section \ref{sec:discussion} we discuss our results 
and in Section \ref{sec-discuss-conclusions} we give our conclusions. 

\section{Set-up of numerical experiments}
\label{sec-setup}

\begin{table}
\scalefont{0.8}
\tabcolsep 3pt
 \caption{Properties of group environments and galaxies.}
 \label{group-param}
 \begin{tabular}{@{}llllr@{}}
  \hline
                            & ``$z$=0''     & ``$z$=1''    &  ``$z$=2''   &                            \\
  \hline
  Group DM halo             &               &              &              &                            \\
  Virial mass               & 9.9           & 9.9          & 9.9          & $(\times 10^{12}M_{\sun}$) \\
  Virial radius             & 555.94        & 329.19       & 226.09       & (kpc)                      \\
  Concentration             & 9.74          & 4.87         & 3.25         &                            \\
  Circular velocity         & 276.97        & 360.07       & 434.62       & (km s$^{-1}$)              \\
  Number particles          & 5.5           & 5.5          & 5.5          & ($\times 10^{5}$)          \\
  Softening                 & 0.55          & 0.32         & 0.22         & (kpc)                      \\
  \hline
  Group stellar spheroid    &               &              &              &                            \\
  Mass                      & 1             & 1            & 1            & ($\times 10^{11}M_{\sun}$) \\
  Scale radius              & 3.24          & 1.91         & 1.31         & (kpc)                      \\
  Number particles          & 2.5           & 2.5          & 2.5          & ($\times 10^{5}$)          \\
  Softening                 & 0.1           & 0.06         & 0.04         & (kpc)                      \\
  \hline
  Galaxy DM halo            &               &              &              &                            \\
  Virial mass               & 2.97-39.6     & 2.47-34.6    & 2.97-9.9     & ($\times 10^{11}M_{\sun}$) \\
  Virial radius             & 172.7-409.6 & 96.2-231.9 & 70.2-104.9 & (kpc)                      \\
  Concentration             & 15.36-10.97   & 7.87-5.58    & 5.12-4.38    &                            \\
  Number particles          & 2.5           & 2.5          & 2.5          & ($\times 10^{5}$)          \\
  Softening                 & 0.35          & 0.26          & 0.19        & (kpc)                      \\
  \hline
  Galaxy stellar disc       &               &              &              &                            \\
  Mass                      & 2.8           & 1.42         & 0.72         & ($\times 10^{10}M_{\sun}$) \\
  Scale-length              & 3.5           & 1.65         & 0.9          & (kpc)                      \\
  Number particles          & 5             & 5            & 5            & ($\times 10^{4}$)          \\
  Softening                 & 0.05          & 0.012        & 0.007        & (kpc)                      \\
  \hline
 \end{tabular}
\end{table}

We have carried out 50 simulations of isolated mergers between a single 
galaxy and a larger main halo, in order to quantify their merger timescales. 
Similarly to \citetalias{boylan-kolchin2008}, our basic strategy is to 
release a single disc galaxy at a time, on a bound orbit, from the virial 
radius of the main halo and study the evolution of its mass content and 
orbital angular momentum as it is dragged towards the centre of the halo. 
Our simulations cover the time span since a galaxy is released from the 
virial radius of the main halo until it is either disrupted or it has 
exhausted its initial orbital angular momentum (see Section \ref{def:merger}). 

Each main halo is modelled as a $N$-body self-consistent multi-component 
system, where the DM halo follows a NFW density profile \citep{navarro1997}:
\begin{equation} \label{nfw}
\rho_{\rm{halo}}(r) = \frac{\rho_{\rm s}}{(r/r_{\rm s})(1+r/r_{\rm s})^2} ,
\end{equation}
where $\rho_{\rm s}$ is a characteristic scale density and $r_{\rm s}$ a
scale radius. Additionally, a spherical stellar component, 
resembling a central galaxy, is located at the halo centre, and its mass 
follows a Hernquist density profile \citep{hernquist1990}:
\begin{equation}
\rho_{\rm *}(r) = \frac{M_{\rm *}}{2\pi} \frac{a_{\rm *}}{r(r+a_{\rm *})^3} ,
\label{bulgeprof}
\end{equation}
where $M_{\rm *}$ is the stellar mass and $a_{\rm *}$ is the scale radius.
Similarly, each galaxy is also modelled as a self-consistent multi-component 
system, formed by a stellar disc embedded in a DM halo. The stellar mass 
in discs follows an exponential density profile:
\begin{equation} \label{diskprof}
\rho_{\rm disc}(R,z) = \frac{M_{\rm disc}}{8\pi R_{\rm D}^2 z_{\rm D}}
\exp\left(-\frac{R}{R_{\rm D}}\right)\ \textrm{sech}^2\left(\frac{z}{2z_{\rm D}}\right) ,
\end{equation}
where  $M_{\rm disc}$ is the disc mass, $R_{\rm D}$ is the exponential
scale-length, and $z_{\rm D}$ is the exponential scale-height. The DM 
halo of galaxies is also assumed to follow a NFW profile. All DM haloes 
in our simulations are initially spherical, do not rotate, and the 
structure of their inner region has been adiabatically contracted to 
account for the growth of the baryonic component 
\citep[see][for details]{villalobos-helmi2008}.

The initial orbital parameters of discs are chosen to be consistent with 
distributions of orbital parameters of infalling substructures, obtained 
from cosmological simulations \citep[e.g.,][]{benson2005,wetzel2011}.
In this paper, we focus on the most likely orbital circularity 
of infalling substructures, and on the extreme values of the distributions.

The initial conditions of our isolated mergers are placed in a simplified 
context of three different redshift epochs, ``$z$=0'', ``$z$=1'' and 
``$z$=2''. In this context, the \emph{initial} structure of DM haloes is 
defined by both the virial over-density $\Delta_{\rm vir}(z)$ \citep{bryan-norman1998} 
and the halo concentration $c(M_{\rm vir},z)$.\footnote{We assume 
$\Omega_{m,0}=0.3$, $\Omega_{\Lambda}=0.7$ and $H_0=70$ km~s$^{-1}$~Mpc$^{-1}$.} 
Note that during the simulations, a main halo \emph{only} interacts with 
a single disc galaxy, and does not accrete additional mass. 
As for the evolution of halo concentration, we adopt 
$c(M_{\rm vir},z) \propto (1+z)^{-1}$ \citep{bullock2001}.

At all three redshifts, main haloes have a mass of 10$^{13} M_{\sun}$. 
This choice is based on the mass range reported by \citet{mcgee2009} and 
\citet{delucia2012}, where significant environmental effects on galaxies 
must take place \citep[see also][]{berrier2009}. We have kept fixed both 
the mass and scale-lengths of stellar discs, while covering a range of 
masses and radii of the DM haloes in which they are embedded (see below).
The stellar-to-DM mass ratios of disc galaxies are consistent with 
stellar-to-DM mass relations from both observations and theoretical studies 
\citep{behroozi2010,guo2010,moster2010}. Table~\ref{group-param} lists the 
structural parameters of main haloes and disc galaxies. 
Table~\ref{list-exper-z0} provides the range of the parameter space  
covered by our experiments. These parameters delimit the validity of our results. 
Note that our experiments do not include cases where 
$c_{\rm host} \sim c_{\rm sat}$ which would be relevant in the 
case of major mergers. 

We have carried out test simulations to ensure that our results are not 
affected by numerical resolution. All simulations were run using 
{\small GADGET-3} \citep{springel2005}.

\begin{table}
\scalefont{0.8}
\tabcolsep 4.3pt 
\begin{minipage}{84mm}
 \caption{Ranges of parameters explored by the experiments.}
 \label{list-exper-z0}
 \begin{tabular}{@{}ccccccc@{}}
  \hline
 $z$       & $M_{\rm host}/M_{\rm sat}$ & $\eta$     & $r_{\rm c}(E)/r_{\rm vir}$ & $t_{\rm DM}$ & $t_{\rm J}$ & $t_{\rm stars}$ \\
 $^{(1)}$  & $^{(2)}$                   & $^{(3)}$   & $^{(4)}$                   &  $^{(5)}$    & $^{(6)}$    & $^{(7)}$        \\
  \hline
 0         & 2.5--33.33                 & 0.21--0.91 &  0.91--1.94                & 4.8--26.5    & 4.8--28.5   & 5.3--29         \\
 1         & 2.86--40                   & 0.21--0.9  &  1.01--1.97                & 3.2--22      & 3.7--26     & 4.2--26         \\
 2         & 10--33.33                  & 0.55       &  1.06--1.16                & 2.5--10      & 3.5--22     & 4--24           \\
  \hline \hline
 \end{tabular}
\\(1) Redshift. (2) Mass ratio of each merger. (3) Initial orbital 
circularity of the satellite, in units of $V_c(r_{\rm vir})$.
(4): Initial orbital energy of the satellite, expressed in terms of 
the radius $r_c$ of a circular orbit with the same orbital energy $E$, 
in units of $r_{\rm vir}$.
(5): merger time computed when only 5 per cent of the DM halo of a 
satellite remains bound, in Gyr.
(6): merger time computed when only 5 per cent of the initial orbital 
angular momentum of a satellite is retained, in Gyr.
(7): merger time computed when only 5 per cent of the stellar content 
of a satellite remains bound, in Gyr.
\end{minipage}
\end{table}

\section{Results}
\label{sec-results}

\subsection{Definition of ``merger''}
\label{def:merger}

As a galaxy orbits within a halo, it gradually loses orbital energy 
and orbital angular momentum due to dynamical friction (mainly) 
against the medium of the larger halo. As its orbit decays towards the 
densest region of the halo, a galaxy might also undergo mass loss via 
tidal stripping, especially at pericentric passages where tidal forces 
are stronger. Previous studies have used several criteria to define 
when a galaxy has ``merged'' with a larger halo. A galaxy is considered 
merged if either: (i) its remaining bound DM mass is 5 per cent of its 
initial value; or, (ii) the remaining orbital angular momentum is 5 per 
cent of its initial value; or, (iii) the remaining bound stellar mass 
is 5 per cent of its initial value. In general, all three definitions 
yield similar merger times, where usually $t_{\rm DM}<t_{\rm J}<t_{\rm stars}$. 
In the rest of the paper, we adopt the definition of merger time as the 
minimum between $t_{\rm J}$ and $t_{\rm stars}$. This corresponds to 
$t_{\rm J}$ in most of our experiments (e.g. see Fig.~1 in 
\citetalias{boylan-kolchin2008}).

\subsection{Comparison to previous models of merger times}

\begin{figure}
\begin{center}
\includegraphics[width=48mm]{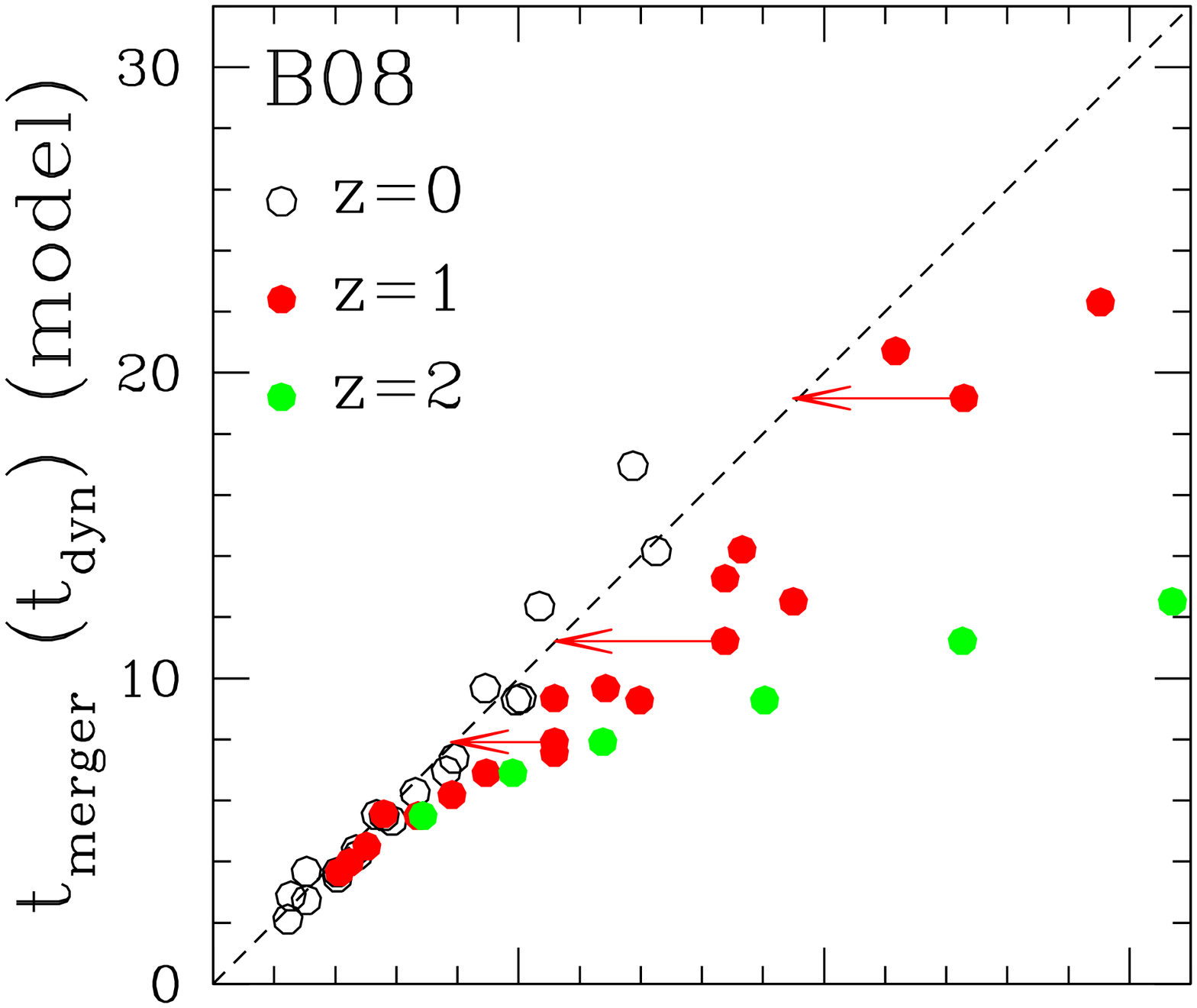}\hspace*{-1.15cm}
\includegraphics[width=48mm]{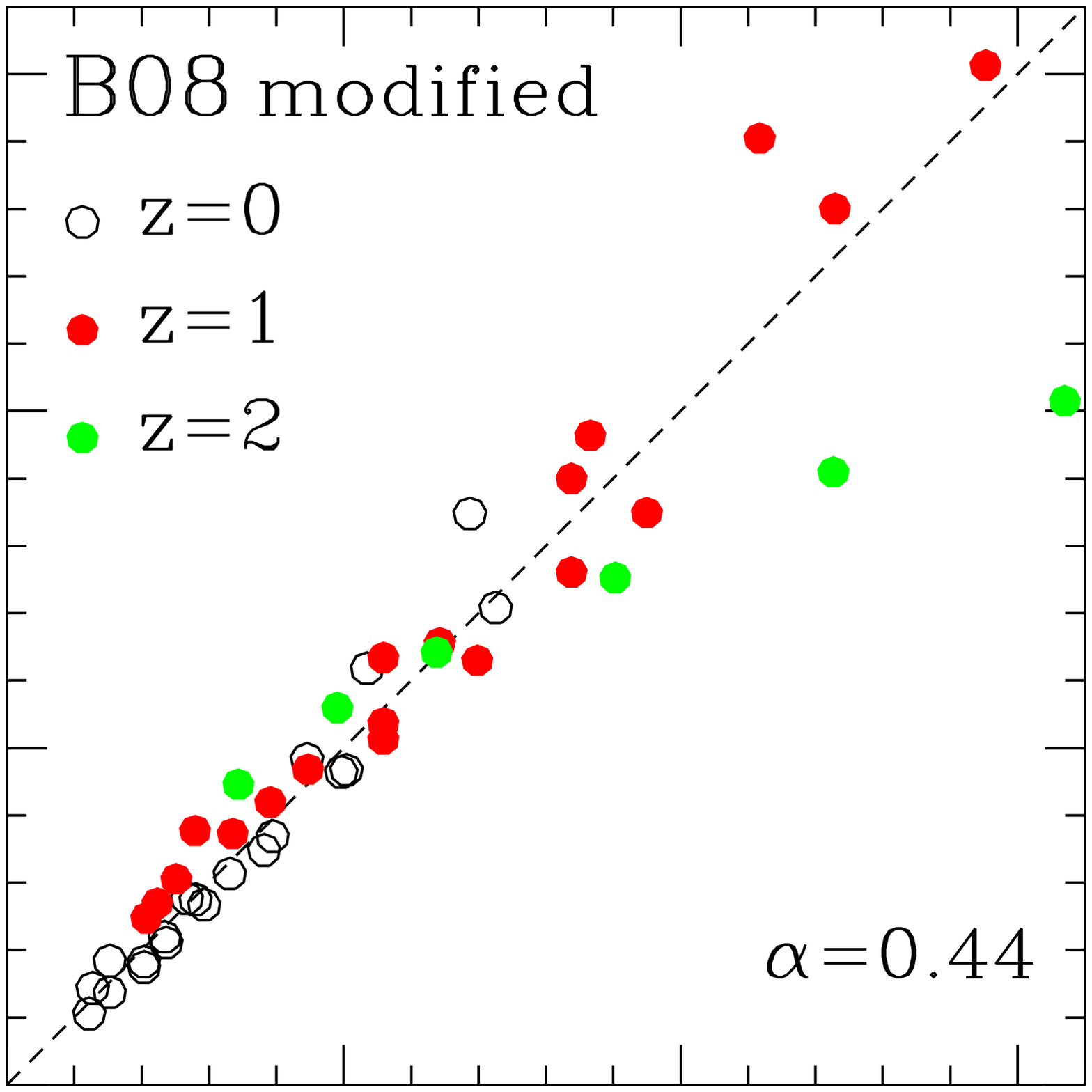}\\\vspace*{-1.11cm}
\includegraphics[width=48mm]{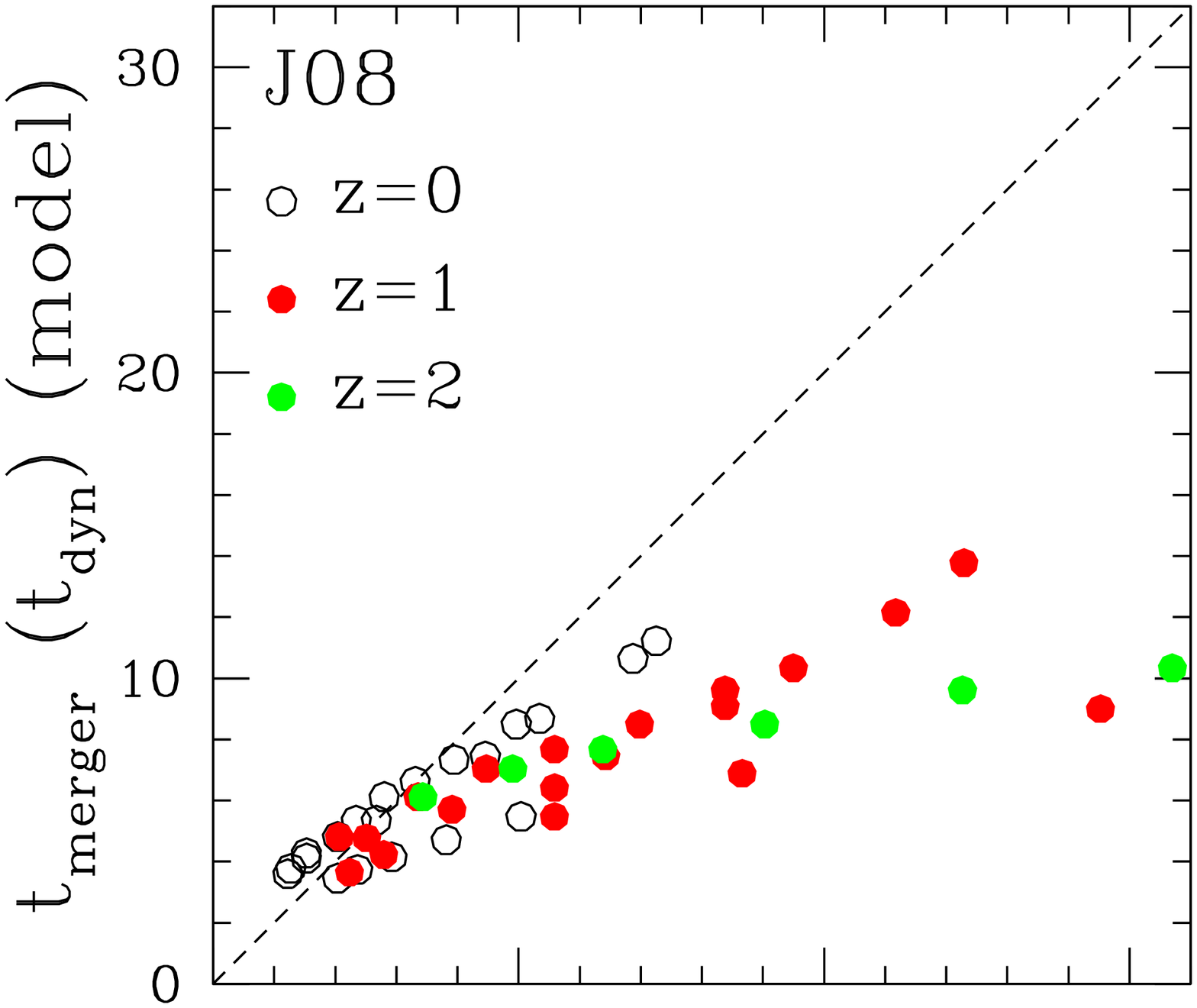}\hspace*{-1.15cm}
\includegraphics[width=48mm]{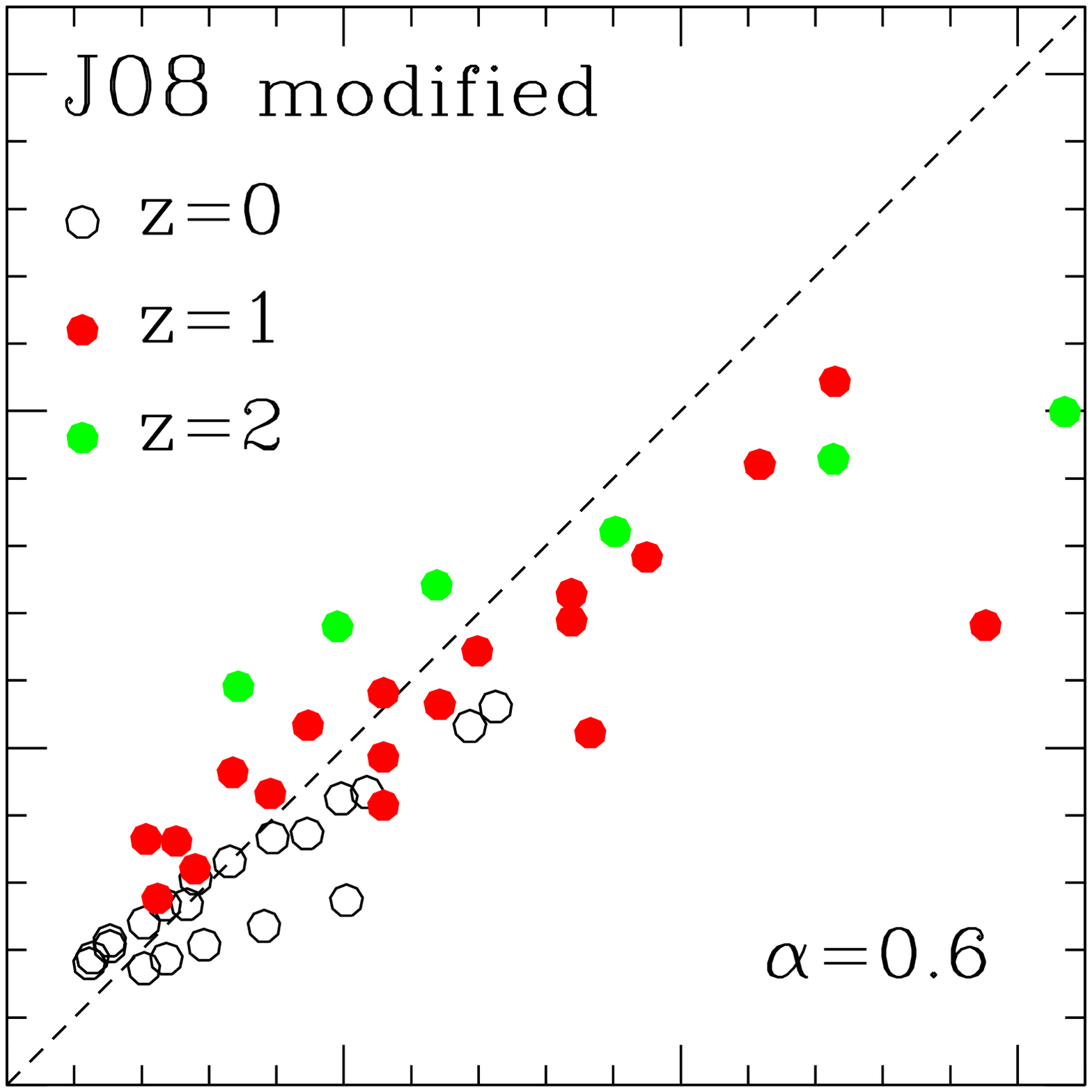}\\\vspace*{-1.11cm}
\includegraphics[width=48mm]{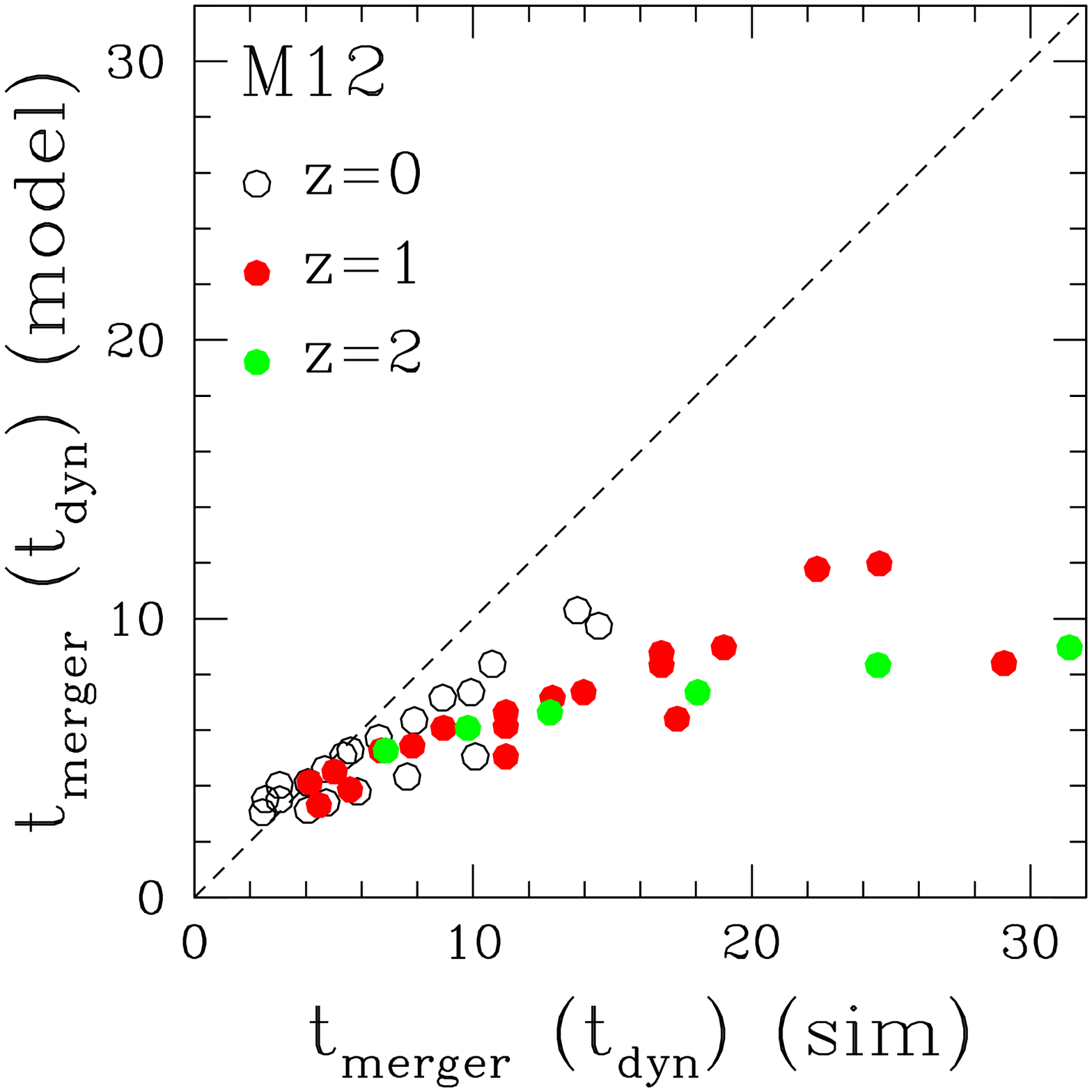}\hspace*{-1.15cm}
\includegraphics[width=48mm]{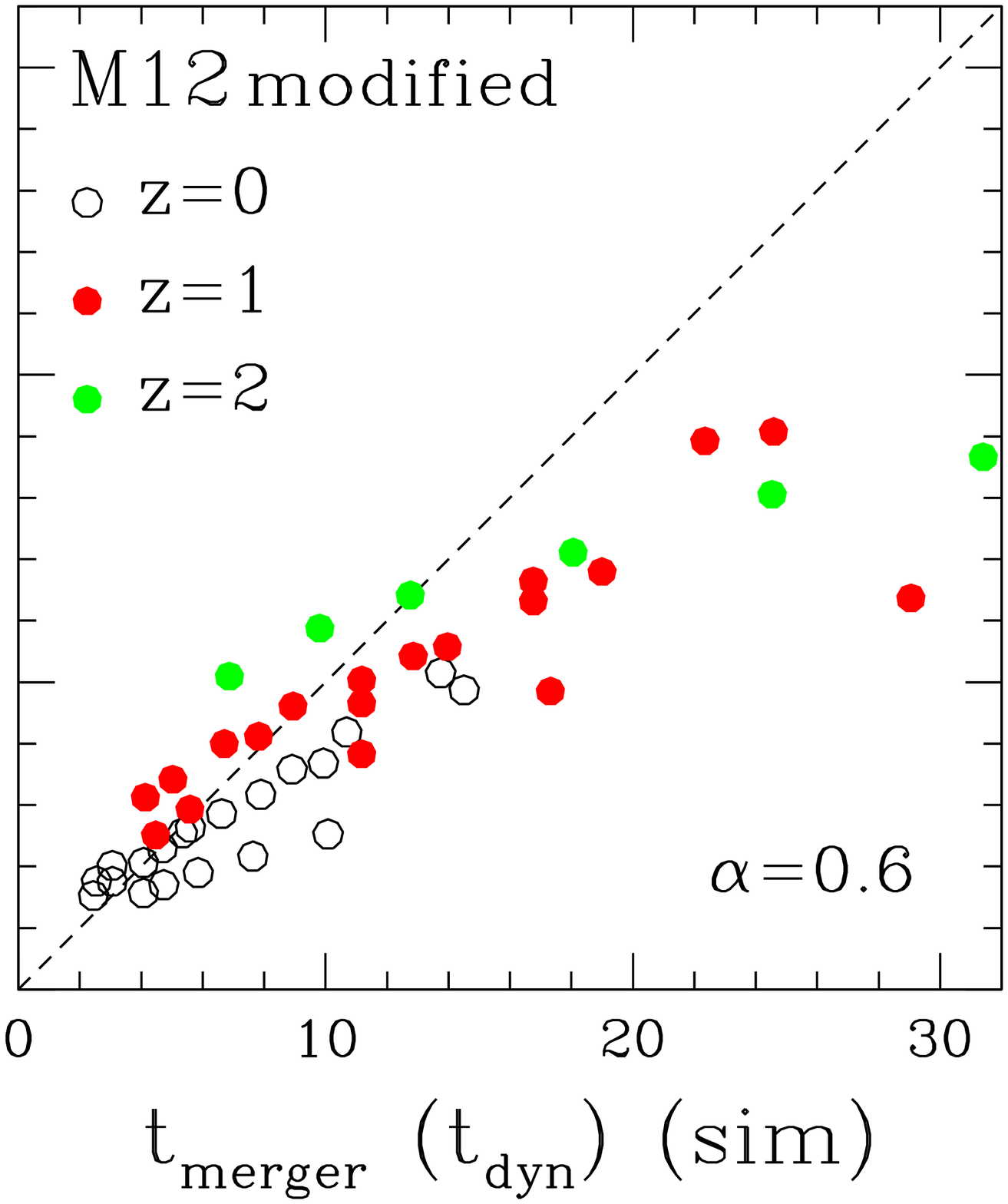}
\end{center}
\caption{Left column: predictions of the \citetalias{boylan-kolchin2008},
\citetalias{jiang2008}, \citetalias{mccavana2012} models applied to our
experiments at ``$z$=0'' (black), ``$z$=1'' (red), and ``$z$=2'' (green).
Arrows indicate the improved accuracy of the prediction for selected
experiments at ``$z$=1'', when mergers are resimulated with a higher initial
concentration of the galaxies' haloes. Right column: predictions of the
respective modified models, $t_{\rm merger}^{\rm mod}(z) = (1+z)^{\alpha} t_{\rm merger}$.}
\label{slope-model}
\end{figure}

We compare the merger times obtained from our simulations to the 
following models of merger timescales from the literature.
\citetalias{boylan-kolchin2008} and \citetalias{mccavana2012}, 
using isolated and cosmological simulations respectively, find:
\begin{equation} \label{bk08-mccav12}
\frac{t_{\rm merger}^{\rm B08}}{t_{\rm dyn}} =
A \frac{(M_{\rm host}/M_{\rm sat})^B}{\ln(1+M_{\rm host}/M_{\rm sat})}  e^{C \eta}  \left[\frac{r_c(E)}{r_{\rm vir}}\right]^D ,
\end{equation}
with the following best fitting parameters 
(A,B,C,D)=(0.216,1.3,1.9,1.0) for \citetalias{boylan-kolchin2008}, and 
(0.9,1.0,0.6,0.1) for \citetalias{mccavana2012}. 
\citetalias{jiang2008}, using cosmological hydro-dynamical simulations, find:
\begin{equation} \label{jiang08}
\frac{t_{\rm merger}}{t_{\rm dyn}} =
\frac{0.94\eta^{0.6}+0.6}{2 \times 0.43} \frac{M_{\rm host}}{M_{\rm sat}} \frac{1}{\ln[1+(M_{\rm host}/M_{\rm sat})]} .
\end{equation}

In all three models, $t_{\rm dyn} = r_{\rm vir}/V_c(r_{\rm vir}) = \sqrt{2/\Delta(z)}/H(z)$, 
where the over-density $\Delta(z)$ and Hubble constant $H(z)$ vary as 
a function of redshift. Note that dynamical timescales do not depend 
on either halo mass or virial radius, but only on the cosmological 
parameters chosen. For the parameters used in this paper, the dynamical 
timescales in our simulations are $t_{\rm dyn}$($z$=0)=1.9644~Gyr, 
$t_{\rm dyn}$($z$=1)=0.895~Gyr, and $t_{\rm dyn}$($z$=2)=0.5095~Gyr. 

\begin{figure}
\begin{center}
\includegraphics[width=75mm]{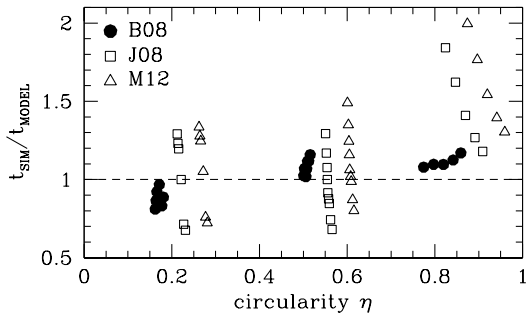}\vspace*{-0.5cm}
\end{center}
\caption{Comparison between predictions of the \citetalias{boylan-kolchin2008}, 
\citetalias{jiang2008} and \citetalias{mccavana2012} models applied to our 
experiments at ``$z$=0'', as a function of the initial orbital circularity 
of satellites. Predictions from \citetalias{boylan-kolchin2008} and 
\citetalias{mccavana2012} are shown with offsets $\eta$-0.05 and $\eta$+0.05, 
respectively, for clarity.
}
\label{comp-models}
\end{figure}

Figure~\ref{slope-model} (left) shows the predicted merger times from 
the \citetalias{boylan-kolchin2008}, \citetalias{jiang2008} and 
\citetalias{mccavana2012} prescriptions applied to our mergers. For 
experiments at ``$z$=0'', \citetalias{boylan-kolchin2008} offer 
accurate predictions at least up to $\sim$15 $t_{\rm dyn}$, 
independently of the merger definition used (Section~\ref{def:merger}). 
An analysis of predicted merger timescales as a function of the 
initial orbit of galaxies shows that the \citetalias{boylan-kolchin2008} 
model tends to slightly overestimate (underestimate) $t_{\rm merger}$ 
for galaxies with lower (higher) initial circularities 
(Figure~\ref{comp-models}). Also, \citetalias{boylan-kolchin2008} 
appears to be fine-tuned to accurately predict merger timescales for 
galaxies infalling with the most likely orbital circularities 
($\eta$$\sim$0.6). The \citetalias{boylan-kolchin2008} model, applied 
to our experiments at ``$z$=0'', exhibits an overall scatter of $\pm$20 
per cent across the range of explored circularities, decreasing to 
$\pm$10 per cent over different mass ratios at a fixed circularity. 
The high accuracy of the \citetalias{boylan-kolchin2008} prescription 
is not surprising, since it was obtained from simulations of isolated 
mergers at $z$=0, similar to those presented in this paper. However, for 
experiments at ``$z$=1'' and ``$z$=2'', a comparable level of accuracy 
in the predictions is only reached up to $\sim$8 $t_{\rm dyn}$ 
(Figure~\ref{slope-model}, top-left). Beyond that limit, 
\citetalias{boylan-kolchin2008} systematically under-predicts merger times. 
Larger differences are observed for experiments with longer merger times 
and for those at higher redshift. As \citetalias{boylan-kolchin2008}, 
\citetalias{jiang2008} and \citetalias{mccavana2012} also significantly 
under-predict longer merger times, and increasingly so for mergers at 
higher redshift when compared to our controlled merger simulations. This 
implies that the implicit redshift dependence in all three models, from 
$t_{\rm dyn}(z)$, is not enough to account for the measured evolution 
of merger timescales.

In general, we find that the \citetalias{jiang2008} and 
\citetalias{mccavana2012} models yield similar predictions for our 
experiments at ``$z$=0'', although showing a significantly larger scatter 
in comparison to the \citetalias{boylan-kolchin2008} model
(Figure~\ref{comp-models}). In spite of the larger scatter, 
\citetalias{jiang2008} and \citetalias{mccavana2012} do predict accurately 
the median $t_{\rm merger}$ (for mergers of different mass ratios) for 
experiments with low and most likely initial orbital circularities.
This is possibly a consequence of the fact that, in both studies, 
mergers are extracted from cosmological simulations, and so include 
a large number of radial mergers. Both models, however, under-predict 
the merger timescales in case of high circularities, which are associated 
to long merger times, showing an offset of $\sim$55 per cent. The scatter 
associated to both \citetalias{jiang2008} and \citetalias{mccavana2012} 
models is $\pm$35 per cent, for mergers with low and most likely initial 
orbital circularities.

We argue that the \citetalias{boylan-kolchin2008}, \citetalias{jiang2008}
and \citetalias{mccavana2012} models under-predict long merger times at 
high redshift as a consequence of both neglecting the evolution of halo 
concentration in satellite, and under-sampling of long merger times. 

\subsubsection{Evolution of halo concentration in satellites}

In order to study the effect of changes in the density profile of satellite 
haloes, we have resimulated three experiments at ``$z$=1'', increasing the 
concentration of their DM haloes. Specifically, we have given to those DM 
haloes a higher concentration as haloes of the same mass would have at 
``$z$=0''. Arrows in Figure~\ref{slope-model} (top-left) show that 
predictions from the \citetalias{boylan-kolchin2008} model become significantly 
more accurate in the case of mergers with more concentrated DM haloes.

A satellite with a higher concentration is expected to be more resilient 
against tidal disruption. Therefore, it will retain more of its mass, 
leading to a shorter merging timescale. Since the 
\citetalias{boylan-kolchin2008} model is obtained from isolated mergers 
at $z$=0, satellite haloes are assumed to have a too high concentration 
when the model is applied to mergers at $z$$>$0, which leads to shorter 
predicted merger times. 

This result strongly suggests that the systematic under-prediction of 
longer merger timescales by the \citetalias{boylan-kolchin2008} model 
comes as a consequence of the evolution of halo concentration in our 
experiments [$c(M_{\rm vir},z) \propto (1+z)^{-1}$, Section~\ref{sec-setup}], 
which is not included in their simulations. Note that the evolution 
of halo concentration also puts a strain on the validity of merger times 
models derived from the Chandrasekhar dynamical friction formula
\citep{chandrasekhar1943}, as they are applied to mergers of \emph{extended} 
objects.

Motivated by the inverse proportionality between $t_{\rm merger}$ and $c$ in our 
experiments, we introduce a modification to the \citetalias{boylan-kolchin2008} 
model in the form \mbox{$t_{\rm merger}^{\rm mod}(z) = (1+z)^{\alpha} t_{\rm merger}^{B08}$.}
We find that the inclusion of this modification, with $\alpha$=0.44,
offers an improvement in the predicted merger timescales at least up to 
$\sim$20~$t_{\rm dyn}$ for mergers between 0$\le$$z$$\le$2 
(Figure~\ref{slope-model}, top-right).\footnote{The time dependence in this modification 
comes only indirectly through the median concentration-redshift relation.}
Note however that a factor $\sim$2 variation in   
concentration at fixed halo mass \citep{neto2007}, 
implies a $\sim$1.4 variation in $t_{\rm merger}$.

In both \citetalias{jiang2008} and \citetalias{mccavana2012}, the 
evolution of halo concentration is already included in their cosmological 
simulations by construction. No details about the concentration of haloes 
are however discussed in these studies. 

\begin{figure}
\begin{center}
\includegraphics[width=42mm]{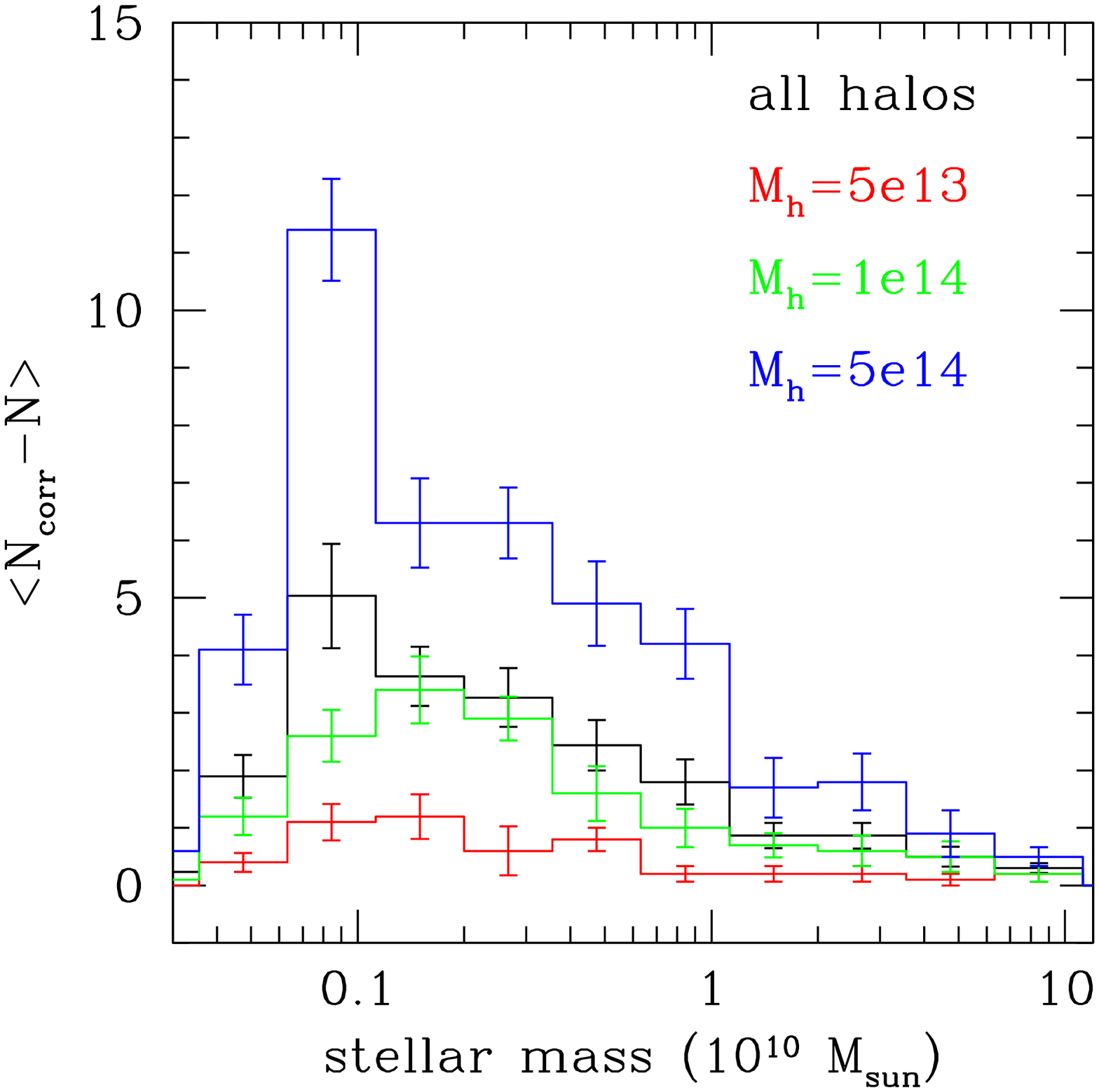}\hspace*{-0.2cm}
\includegraphics[width=42mm]{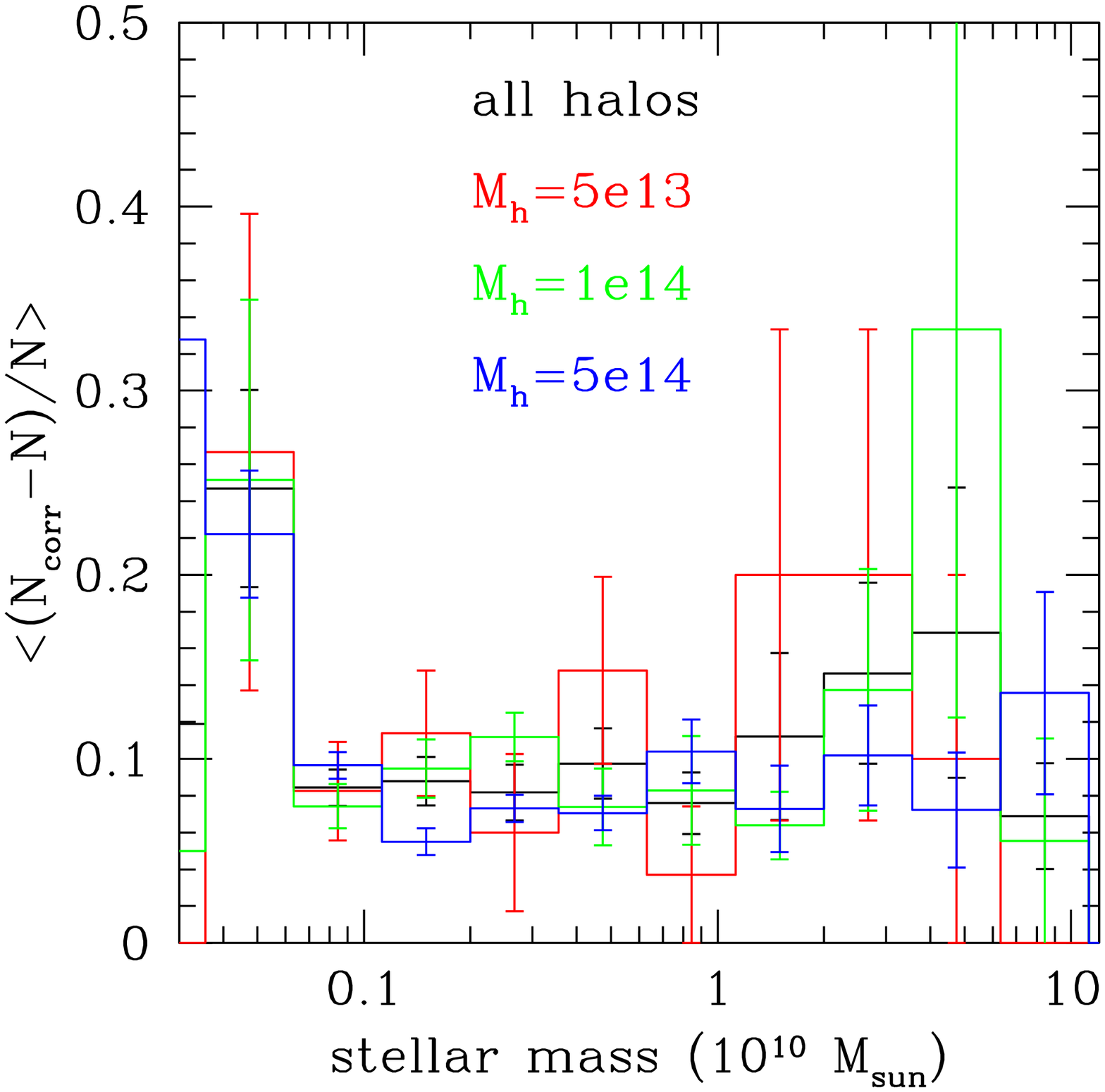}\vspace*{-0.1cm}
\end{center}
\caption{
Left: difference between galaxy stellar mass functions using
$t_{\rm merger}^{\rm mod}(z)$ and $t_{\rm merger}^{\rm B08}$ models, averaged
over 10 haloes in each mass range: $5 \times 10^{13} M_{\sun}$, $10^{14} M_{\sun}$
and $5 \times 10^{14} M_{\sun}$ at $z$=0. Right: averaged fractional increase
(in each galaxy stellar mass bin) associated to $t_{\rm merger}^{\rm mod}(z)$,
with respect to the $t_{\rm merger}^{\rm B08}$ model.
}
\label{slope-model-smf}
\end{figure}

\subsubsection{Under-sampling of long merger times}

Even though cosmological simulations are statistically robust and provide 
a realistic context for studies of merger timescales, in general both long 
timescales ($>$$15~t_{\rm dyn}$) and mergers with high circularities are 
severely under-represented. For instance, only mergers completed by $z$=0 
and with orbital pericentres $\leq r_{\rm vir}$ are considered in 
\citetalias{mccavana2012}. As a consequence, this likely leads to 
inaccurate predictions of long merger timescales.
Even though the motivation behind our modification does not apply 
to the \citetalias{jiang2008} and \citetalias{mccavana2012} models (since 
evolution of halo concentration is included in their models by construction), our proposed
modification does partially alleviate the disagreement between their predictions
and our simulations, as shown in Figure~\ref{slope-model}.
Note that \citetalias{boylan-kolchin2008} might also be affected by under-sampling 
given the relatively low number of long mergers their fitting formula is based on.

\section{Discussion}
\label{sec:discussion}

We study the effect of our modification to the \citetalias{boylan-kolchin2008} 
model on the galaxy stellar mass functions for galaxies residing in haloes 
of different mass.

The galaxy stellar mass functions are computed by selecting from the 
Millennium simulation \citep{springel2005b} 30 haloes in three mass 
ranges ($5 \times 10^{13}$ $M_{\sun}$, $10^{14}$ $M_{\sun}$, and 
$5 \times 10^{14}$ $M_{\sun}$ at $z$=0). Following their main progenitors 
back in time, we extract all information about the galaxies residing in 
each halo and in its main progenitors, taking advantage of publicly 
available galaxy catalogues \citep{delucia2007}. For each galaxy, we then 
store both its mass and that of its host halo at accretion time (i.e. the 
snapshot before it becomes a satellite for the first time). In the 
procedure, we take care of avoiding counting the same galaxy more than 
once. An orbital circularity is assigned randomly to each galaxy from 
distributions of orbital parameters obtained by \citet{benson2005}. 
Then, both $t_{\rm merger}^{\rm B08}$ and $t_{\rm merger}^{\rm mod}(z)$  
prescriptions are applied to each galaxy and its host. Finally, it is 
assumed that a galaxy has survived by $z$=0, if 
$t_{\rm merger}$$>$$t_{\rm LB}(z_{\rm accretion})$, where $t_{\rm LB}$ 
is the look-back time at the redshift of accretion. As a first 
approximation, we assume that all surviving galaxies conserve their 
stellar mass since they were accreted. We also assume that the stellar 
mass of galaxies that do not survive is added to either the diffuse stellar 
component of the main halo, or to the central galaxy. This is clearly a 
strong simplification, since mergers between satellite galaxies within a 
main halo could also take place, altering the intermediate-mass region of 
the mass functions. We plan to study in a future work the consequences of 
the proposed modified model in the context of a more realistic galaxy 
formation model.

Figure~\ref{slope-model-smf} (left) shows the difference between mass 
functions using $t_{\rm merger}^{\rm mod}(z)$ and $t_{\rm merger}^{\rm B08}$ 
models, averaged over 10 haloes in each mass range. We find that the 
modification introduced leads to a larger number of lower mass satellites 
surviving at $z$=0, across all explored mass ranges. This comes as a 
natural consequence of the under-prediction of long merger timescales 
(associated to low mass satellites) at high redshift by the 
\citetalias{boylan-kolchin2008} model, as seen in 
Figure~\ref{slope-model} (left).

Figure~\ref{slope-model-smf} (right) shows the fractional increase in 
the number of satellites within each galaxy stellar mass bin, due to the 
use of $t_{\rm merger}^{\rm mod}(z)$ with respect to the $t_{\rm merger}^{\rm B08}$ 
model. We find that satellites on the low mass end present the most 
significant fractional increase: $\sim$25 per cent considering haloes in 
all mass ranges. On the other hand, more massive satellites show a rather 
constant (within the standard errors) fractional increase of $\sim$10 per 
cent. Previous studies comparing observational data to predictions from 
semi-analytic models of galaxy evolution have found that these over-predict 
the number of faint galaxies \citep{weinmann2006b,liu2010}. Our results 
suggest that this problem would worsen if more realistic models of merger 
timescales are employed. 

By assuming that all the stellar content of merged galaxies contributes 
to the mass of central galaxies, we find that $t_{\rm merger}^{\rm mod}(z)$ 
leads to a mass reduction of $\sim$10 per cent for the central galaxy, 
with respect to $t_{\rm merger}^{\rm B08}$. This is because there is less 
contribution from merged satellites, which would have longer survival 
times. This calculation assumes that the initial mass of the central 
galaxy (i.e., before mergers take place) is 1 per cent that of the final 
halo mass at $z$=0.  

Similar results are found when we use modified versions of both the
\citetalias{jiang2008} and \citetalias{mccavana2012} models instead of 
\citetalias{boylan-kolchin2008}.

\section{Summary and Conclusions}
\label{sec-discuss-conclusions}

In this work, we compare predictions from three models of merger times 
available in the literature to simulations of isolated mergers.
We generate 50 simulations of single mergers between a satellite galaxy 
and a main halo at three redshift epochs, studying the evolution of the 
galaxy mass content and orbital angular momentum, as it is affected by 
dynamical friction. We probe a parameter space of different halo 
concentrations, merger mass ratios, orbital circularities, and orbital 
energies of galaxies.
 
We find that the implicit dependency on redshift in the models is not 
enough to account for variations as a function of redshift in our 
simulations. In particular, we find that prescriptions available in the 
literature significantly under-predict long timescales for mergers at 
high redshift. In a prescription derived from simulations of isolated 
mergers, this is found to be caused mainly by the lack of an explicit 
treatment of the evolution of halo concentration in satellite galaxies.
On the other hand, in prescriptions derived from cosmological simulations, 
the disagreement is likely due to the fact that long merger times, as well 
as mergers with high initial orbital circularities, are under-represented.

Motivated by the effect of the evolution of halo concentration of 
satellites, we introduce a modification to the model derived
from isolated mergers in the form $t_{\rm merger}^{\rm mod}(z) = (1+z)^{\alpha} t_{\rm merger}$. 
With $\alpha$=0.44, the prescription improves significantly the 
predictions of merger times up to $\sim$20~$t_{\rm dyn}$ for mergers 
between 0$\le$$z$$\le$2. We estimate that our proposed modification can 
lead up to a 25 per cent increase in the number of low mass galaxies in 
massive haloes, and a 10 per cent increase in the number of the most 
massive galaxies. This would worsen the disagreement between observations 
and predictions from semi-analytic models of galaxy evolution, as found 
in previous studies.  

Precise predictions of merger timescales are a key ingredient in models 
of galaxy evolution. In a future work, we plan to investigate in detail 
the influence of our proposed modification, in the context of a realistic 
galaxy formation model coupled to $N$-body cosmological simulations.

\section*{Acknowledgements}

\'AV and GDL acknowledge funding from ERC grant agreement n. 202781. 
SW acknowledges funding from ERC grant HIGHZ no. 227749. All simulations 
were run in CASPUR HPC facilities. The authors 
thank Volker Springel for making available the non-public version of the 
{\small GADGET-3} code. This work has been partially supported by the 
Marie Curie Initial Training Network CosmoComp (PITN-GA-2009-238356), 
the PRIN-INAF09 project ``Towards an Italian Network for Computational Cosmology'', 
the PRIN-MIUR09 ``Tracing the growth of structures in the Universe'' 
and the PD51 INFN grant.

\bibliographystyle{mn.bst}
\bibliography{mn-jour,biblio}

\bsp

\label{lastpage}

\end{document}